\documentclass[aps,pra,showpacs,twocolumn, longbibliography]{revtex4-2}

\usepackage{graphicx}
\usepackage{dcolumn}
\usepackage{bm}
\usepackage{color}
\usepackage{amsmath}
\usepackage{amssymb}
\usepackage{comment}
\usepackage{xcolor}

\usepackage[normalem]{ulem}

\DeclareRobustCommand{\erase}{\bgroup\markoverwith{\textcolor{red}{\rule[.5ex]{2pt}{0.4pt}}}\ULon}
 
\begin{document}


\title{Experimental investigation of single qubit quantum classifier with small number of samples}

\author{Shunsuke Abe$^{1}$}

\author{Shota Tateishi$^{1}$}

\author{Roga Wojciech$^{2}$}

\author{Masahiro Takeoka$^{2}$}

\author{Takafumi Ono$^{1}$}
 \email{ono.takafumi@kagawa-u.ac.jp}

\affiliation{%
~\\
$^{1}$Program in Advanced Materials Science
Faculty of Engineering and Design,
Kagawa University,
2217-20 Hayashi-cho, Takamatsu, Kagawa
761-0396, Japan
}
\affiliation{%
~\\
$^{2}$Department of Electronics and Electrical Engineering, Keio University,
3-14-1 Hiyoshi, Kohoku-ku, Yokohama 223-8522, Japan
}

\date{\today}

\begin{abstract}
We experimentally investigated a single-qubit quantum classifier implemented on a silicon photonic integrated circuit, focusing on its performance under photon-limited conditions. Using the Data Reuploading method with layer-wise optimization via Sequential Minimal Optimization (SMO), input data were encoded into the photonic circuit, and classification was performed based on output detection probabilities. Heralded single photons, generated via spontaneous four-wave mixing in a silicon waveguide, served as the input states. Even when the average number of photon samples per input was reduced to approximately two, the classifier achieved nearly 90\% accuracy, provided that the training dataset was sufficiently large. The experimental results were consistent with numerical simulations, which also indicated that performance at low sample sizes can be improved by increasing the size of the training dataset. These findings demonstrate that photonic quantum classifiers can operate effectively with very few photons, supporting their practical feasibility for resource-efficient quantum machine learning on integrated photonic platforms.
\end{abstract}

\maketitle
\section{Introduction}
A classifier is an algorithm that categorizes input data into predefined classes based on its features \cite{Allan1977,Lin2018,Li2021,Cong2022,Mourgias-Alexandris2022}. By analyzing these features, it infers the most probable category for each input. Classifiers are widely used in various domains such as medical diagnosis \cite{Ahsan2022,Alizadehsani2019}, traffic forecasting \cite{Afandizadeh2024,9352246}, and natural language processing \cite{Gasparetto2022,Li2022}. Recently, quantum classifiers—which leverage the principles of quantum mechanics—have gained significant attention \cite{Rebentrost2014,McClean2016,Mitarai2018,Stokes2020,Blank2020,Schuld2020,LaRose2020,Cerezo2021,Xu2021}. These models seek to surpass classical approaches by exploiting quantum properties to enhance computational capabilities and efficiency \cite{Dong2008,PhysRevLett.114.110504,Krizhevsky2017,Fosel2018,Killoran2019,Benedetti2019,Chen2020,Mangini2020,Yunchao2021,Zhang2021b,Chabaud2021,Chen2021a,Nakaji2022,Sen2022,Cimini2023,Bowie2024,Shin2024,Mandilara2024,Roncallo2025}

A typical quantum classification procedure involves three key stages: quantum state preparation, data encoding via quantum operations, and quantum measurement. Training such a classifier entails embedding classical data into a quantum circuit and optimizing the circuit parameters so that the output matches the desired labels. This is typically achieved by minimizing a cost function defined over the measurement outcomes. The performance of quantum classifiers depends crucially on the quantum model and circuit architecture employed, with various platforms having been explored, including trapped ions \cite{Sonika2021,PhysRevA.111.052436}, photonic systems \cite{Bartkiewicz2020, Wang:21}, and superconducting qubits \cite{Havlicek2019,PhysRevResearch.4.033190}.

In this work, we focus on the Data Reuploading method—a recently proposed approach that we have experimentally implemented on a silicon photonic integrated circuit \cite{PerezSalinas2020,PhysRevA.106.012411,PhysRevLett.131.013601}. This method enables the construction of a universal quantum classifier using photons. The Data Reuploading model is structurally similar to neural networks, allowing classical machine learning strategies to be adapted to the quantum setting \cite{Zuo2019,Cong2019,Sui2020,Jerbi2021}. A fundamental distinction between classical and quantum classifiers, however, lies in the probabilistic nature of quantum measurements. While classical models (often based on continuous light sources) yield deterministic outputs, quantum classifiers—particularly those based on single photons—exhibit intrinsic statistical fluctuations due to the discrete nature of quantum measurement. These quantum fluctuations introduce probabilistic errors that can complicate the training process.

Theoretical studies have suggested that accurate training can still be achieved even with extremely limited photon resources—down to a single photon per measurement—provided that the number of training samples is sufficiently large \cite{Sweke2020}. However, experimental validation is crucial to determine the practical viability of such models.

In this study, we implemented a single-qubit quantum classifier using a silicon photonic integrated circuit and experimentally evaluated its performance under photon-limited conditions. Specifically, we reduced the number of photons used during training to levels where statistical fluctuations significantly affected the output. Building on a previous method \cite{Roga2023}, we evaluated the accuracy of cost function estimation within the Data Reuploading framework under such constraints. Additionally, we performed numerical simulations to investigate how classifier performance varies with sample size and training set size across specific classification tasks. Finally, we experimentally realized the classifier using heralded single photons generated on-chip, achieving classification with an average of approximately two photons per training input. Remarkably, we attained a classification accuracy of around 90\%, demonstrating that effective training is possible even under severely photon-constrained conditions. These results represent a meaningful step toward the practical implementation of quantum machine learning technologies.

\section{Theory}
\begin{figure}[t]
 \centering
 \includegraphics[width=\columnwidth]{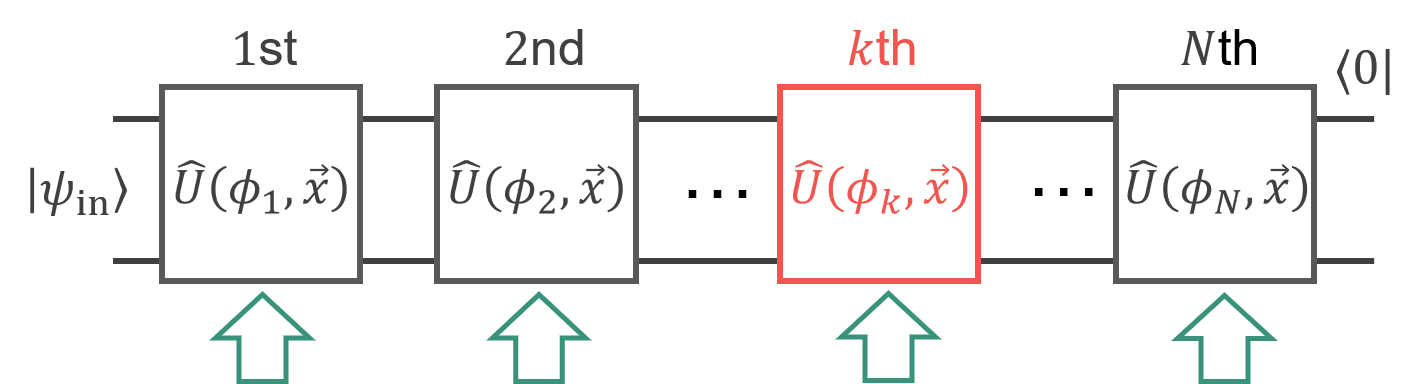}
 \caption{The implementation model of a quantum classifier employing the Data Reuploading method. The input state $|\psi_{\rm{in}}\rangle$ is represented as a two-mode single-photon state. The unitary transformation within the circuit is segmented into $N$ layers. The boxes depicted in the figure represent the respective unitary transformations.
}
 \label{fig1}
\end{figure}

Figure~\ref{fig1} illustrates the architecture of our single-qubit quantum classifier based on the Data Reuploading method. The system encodes a qubit using two spatial modes of a single photon: the upper mode represents the logical state $|0\rangle$, and the lower mode represents $|1\rangle$. The quantum circuit applies a sequence of unitary transformations—implemented using beam splitters and phase shifters—which correspond to rotations of the qubit state on the Bloch sphere.

In the Data Reuploading approach, the quantum circuit is divided into multiple layers. At each layer, the input data vector $\vec{x}$ is re-encoded into the circuit via a phase shift $\varphi_k(\vec{\theta}_k, \vec{x})$, where $\vec{\theta}k$ is a set of trainable parameters optimized during training. The overall transformation applied to the initial state $|0\rangle$ is given by:
\begin{equation}
|\psi{\mathrm{out}}\rangle = \hat{U}(\varphi_1) \hat{U}(\varphi_2) \dots \hat{U}(\varphi_N) |0\rangle.
\end{equation}
The classifier assigns a label based on the measurement outcome: if the probability of obtaining $|0\rangle$ exceeds a predefined threshold, the input is classified as “yes”; otherwise, it is classified as “no.”

To train the circuit, a set of $N$ labeled training data points ${ \vec{x}^{(i)}, y^{(i)} }$ is provided. The goal is to optimize the parameters $\vec{\theta}_k$ such that the estimated output probability $\tilde{p}^{(i)}(\vec{\theta}_k)$ approximates the corresponding label $y^{(i)}$. We adopt a layer-wise analytical optimization strategy based on the Sequential Minimal Optimization (SMO) method \cite{PhysRevResearch.2.043158,PhysRevLett.131.013601,Roga2023}. The cost function used for training each layer is the mean squared error:
\begin{equation} \label{cost}
\tilde{C}(\vec{\theta}_k) = \frac{1}{N} \sum_{i=1}^{N} \left( \tilde{p}^{(i)}(\vec{\theta}_k) - y^{(i)} \right)^2.
\end{equation}

Following the method introduced in~\cite{Roga2023}, the output probability $\tilde{p}^{(i)}(\varphi_k)$ can be reconstructed using three specific phase settings $\phi = 0, \pm 2\pi/3$, exploiting the trigonometric structure of the quantum circuit:
\begin{equation}
\tilde{p}^{(i)}(\phi_k) = \tilde{A}^{i}_0 + \tilde{A}^{i}_1 \cos \phi_k + \tilde{A}^{i}_2 \sin \phi_k.
\end{equation}
The coefficients $\tilde{A}^{i}_j$ can be determined from the three measurement results using the following linear transformation:
\[
\begin{pmatrix} \tilde{A}^{i}_1 \\ \tilde{A}^{i}_0 \\ \tilde{A}^{i}_2 \end{pmatrix} =
\begin{pmatrix}
\cos(2\pi/3) & 1 & \cos(2\pi/3) \\
0.5 & 0.5 & 0.5 \\
-\sin(2\pi/3) & 1 & -\sin(2\pi/3)
\end{pmatrix}
\begin{pmatrix}
\tilde{p}^{i}(2\pi/3) \\
\tilde{p}^{i}(0) \\
\tilde{p}^{i}(-2\pi/3)
\end{pmatrix}.
\]
When a large number of photon samples is used to estimate each $\tilde{p}^{(i)}$, statistical fluctuations become negligible, and the cost function $\tilde{C}$ closely approximates its true value. However, with a finite number of photons, statistical noise becomes significant and introduces variance in the estimated cost function.

In the extreme case where only one photon is used per measurement, the variance in the estimated cost function can be expressed as:
\begin{equation}
\Delta C(\vec{\theta}_k)^2 = \frac{(\Delta \bar{C}^{(i)})^2}{N},
\end{equation}
where $\Delta \bar{C}^{(i)}$ is the average variance across all $N$ training points. For general sample size $M$ per data point, the variance reduces to:
\begin{equation}
\Delta C(\vec{\theta}_k)^2 = \frac{(\Delta \bar{C}^{(i)})^2}{NM}.
\end{equation}
This relation shows that, even under limited photon resources (i.e., small $M$), the variance in the cost function can be suppressed by increasing the number of training samples $N$. Therefore, achieving reliable training under photon-limited conditions requires balancing between quantum sampling resources and classical data volume—a key insight for practical implementations of photonic quantum classifiers.

\begin{figure}[t]
 \centering
 \includegraphics[width=\columnwidth]{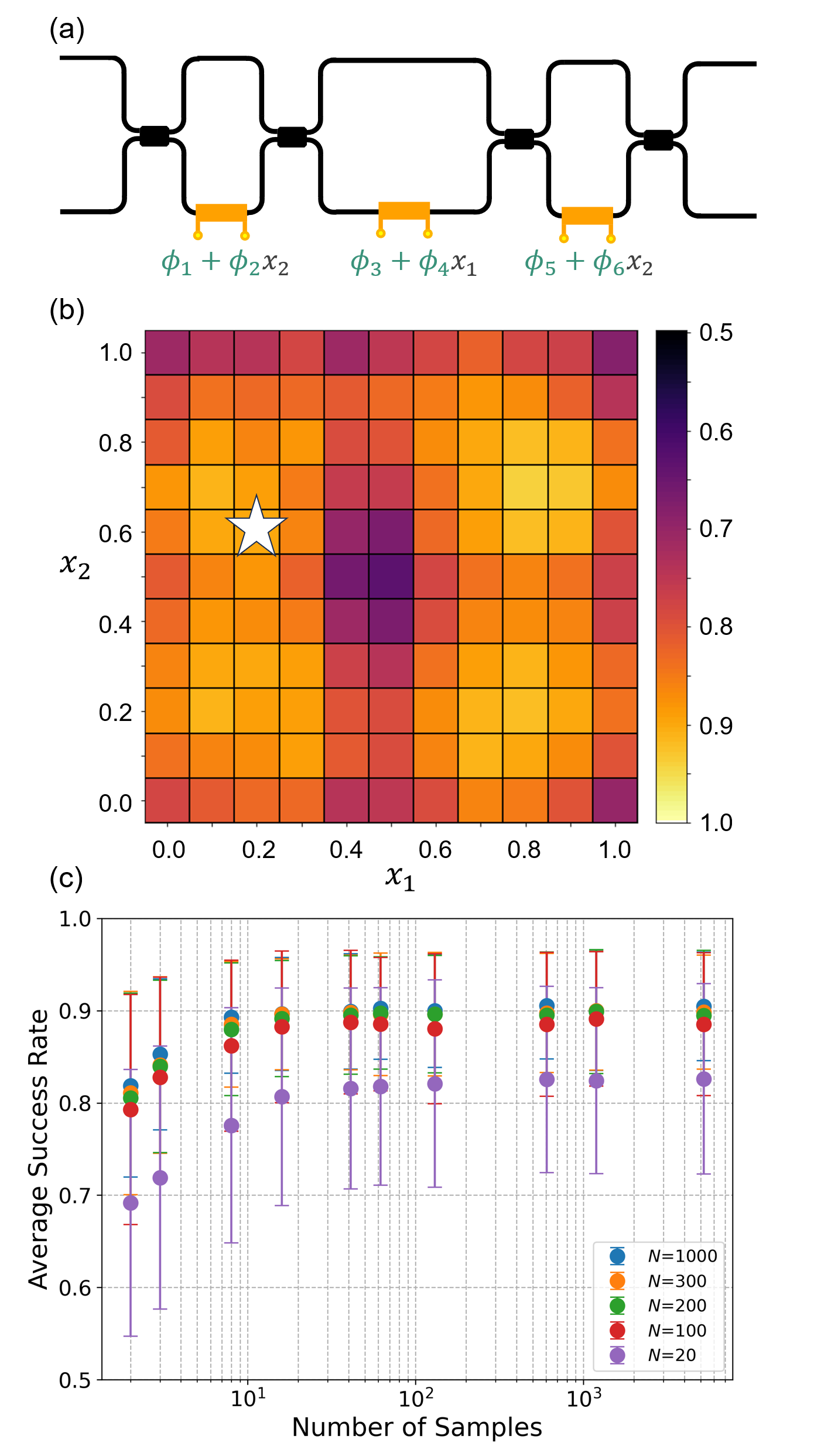}
 \caption{Simulation results for the quantum classifier. (a) Model of the photonic quantum circuit used in the simulation. (b) Dependence of the classification accuracy on the center coordinates of the decision boundary. The simulations and experiments were conducted under the conditions indicated by the star symbol. (c) Dependence of the classification accuracy on the number of photon samples used during training. 
 The error bars represent the variance calculated from 1000 simulation runs.
}
 \label{fig2}
\end{figure}

\section{Simulation results}
\label{sec:simulation}
To evaluate the fundamental performance of the quantum classifier, we conducted numerical simulations based on the Data Reuploading model. Our aim was to clarify how classification accuracy depends on two key parameters: the number of photon samples used to estimate output probabilities and the number of training data points.

Figure~\ref{fig2}(a) shows the optical circuit used in the simulation. It consists of multiple beam splitters and three phase shifters, forming a three-layer structure. Two-dimensional input vectors $\vec{x} = (x_1, x_2)$ were encoded into the phase shifters as illustrated. The circuit contains six tunable parameters $\vec{\phi}$, which were optimized using training data. The classification task involved a circular decision boundary centered at $(X_1, X_2)$ with radius $R$, defined by:
\begin{equation}
(x_1 - X_1)^2 + (x_2 - X_2)^2 = R^2,
\end{equation}
where data points outside the circle were labeled as “yes” and those inside as “no.”

We first trained the classifier under ideal conditions using analytically calculated output probabilities, effectively simulating infinite photon samples to eliminate statistical noise. The radius was set to $R = 0.33$, and 200 training data points were used. Parameter optimization was performed layer by layer using the Sequential Minimal Optimization (SMO) method. The training and evaluation process was repeated 100 times to compute the average classification accuracy.

Figure~\ref{fig2}(b) shows how the average classification accuracy varies with the center position $(X_1, X_2)$ of the decision boundary. Although the Data Reuploading model is theoretically universal, the limitation to three layers introduced some dependence on the boundary's position. Therefore, for consistency, we fixed the boundary parameters for all subsequent simulations and experiments to $R = 0.33$ and $(X_1, X_2) = (0.2, 0.6)$.

We then examined classifier performance under photon-limited conditions. To simulate measurement noise, we assumed that the number of detected photons $M$ follows a Poisson distribution with mean $pM$, where $p$ is the ideal detection probability. Photon counts were numerically generated based on this assumption. Figure~\ref{fig2}(c) shows the average classification accuracy as a function of $M$ for various training data sizes ($N = 20, 100, 200, 300, 1000$). For each $(N, M)$ pair, we performed 1000 independent training trials and averaged the results.

As shown in Fig.~\ref{fig2}(c), classification accuracy increases with the number of photon samples and approaches the ideal value. Remarkably, even with as few as $M \approx 2$ photons per input, the classifier achieves high accuracy when the training dataset is sufficiently large. This supports the theoretical expectation that increased training data can compensate for photon shot noise.

These simulation results highlight the trade-off between quantum sampling efficiency and classical data volume. They also provide a useful benchmark for interpreting experimental data, demonstrating that Data Reuploading-based quantum classifiers remain robust in resource-constrained conditions.

In our simulations, the training dataset size was increased up to 1000 data points. Expanding beyond this (e.g., to several thousand) was not attempted due to computational limitations. Additionally, we observed a noticeable drop in accuracy when $M$ dropped below 10. Whether this degradation can be overcome by further increasing the training dataset remains an open question for future work.

\section{Experimental implementation}
We experimentally implemented a single-qubit quantum classifier on a silicon photonic integrated circuit and evaluated its performance under photon-limited conditions, where the number of photon samples available for training was intentionally restricted. Figure~\ref{fig4}(a) provides an overview of the experimental setup, including the light sources, quantum circuit, and detection system. The chip was fabricated using a commercially available MPW (Multi-Project Wafer) service. The photonic circuit consists of a 4×4 universal unitary transformation circuit, and in this experiment, we used two of its spatial modes to implement the classifier.

\begin{figure}[t]
 \centering
 \includegraphics[width=\columnwidth]{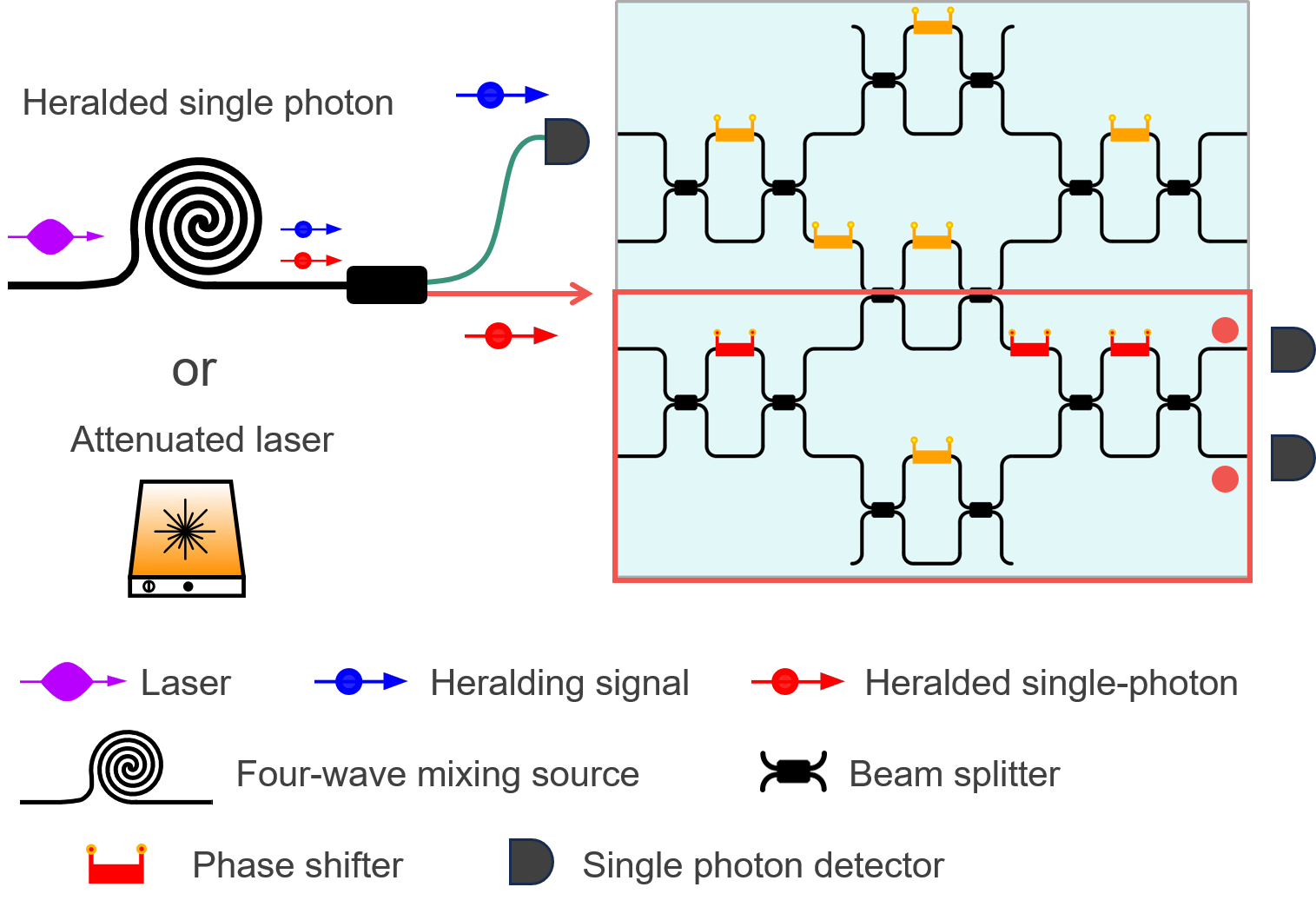}
 \caption{Experimental setup of the single-photon quantum classifier implemented on a silicon photonic integrated circuit. The integrated circuit includes a multi-stage interferometric structure. The circuit enclosed in the red box was used to perform the experiment. Phase shifters used for encoding input data and performing classification are highlighted in red.
}
 \label{fig3}
\end{figure}

Heralded single photons generated via spontaneous four-wave mixing (SFWM) in a silicon waveguide were used as the single-photon input states. In the SFWM process, correlated photon pairs (signal and idler) are produced; detection of the signal photon serves as a herald, indicating the presence of its twin idler photon, which is then injected into the photonic circuit. This heralding enables precise timing of the input photon. A continuous-wave (CW) laser with a power of approximately 10 mW and a center wavelength of 1549.3 nm was used as the pump source. The pump light was coupled into the waveguide via an on-chip spot-size converter, with a coupling loss of about 4 dB, corresponding to approximately 5 mW of optical power coupled into the waveguide. In the experiment, the photon with a central wavelength of approximately 1559.0 nm $\pm$ 0.8 nm was used as the heralding signal, and the photon at 1539.8 nm $\pm$ 0.8 nm was used as the heralded single photon. Under these conditions, the observed generation rate of heralded single photons was approximately 30 counts per second.

Using this photonic circuit and heralded single-photon source, we performed full training and testing in the quantum domain. The optical circuit was structurally identical to that used in the simulation (see Fig.~\ref{fig2}(a)), consisting of a cascaded interferometric network of beam splitters and thermo-optic phase shifters. The classifier implemented a three-layer Data Reuploading model, with each layer applying a unitary transformation controlled by data-encoded phase shifts. Six tunable parameters, $\vec{\phi} = (\phi_1, \phi_2, \phi_3, \phi_4, \phi_5, \phi_6)$, were optimized using the Sequential Minimal Optimization (SMO) algorithm. Two-dimensional input vectors $\vec{x} = (x_1, x_2)$ were mapped to phase shifts at each layer, enabling flexible data encoding within a fixed circuit structure.

\begin{figure}[t]
 \centering
 \includegraphics[width=\columnwidth]{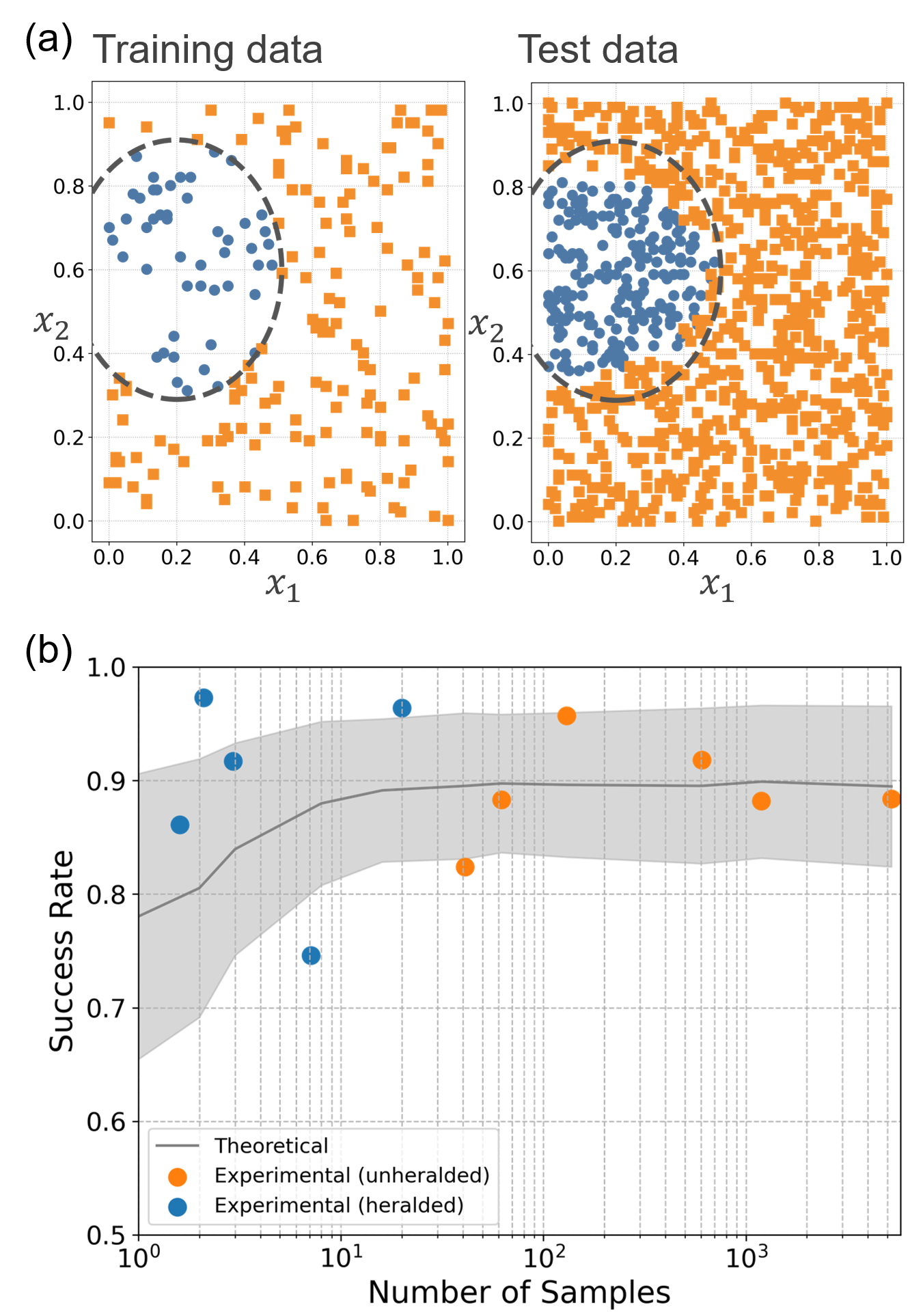}
 \caption{(a) Classification results obtained using heralded single photons. The dashed line represents the boundary of the classification. For the test data, approximately 1000 photon samples were used per input to evaluate the classification accuracy. (b) Dependence of the classification accuracy on the number of photon samples, $M$. The orange dots represent the classification results obtained using pseudo-single photons, where laser light was attenuated to the single-photon level. The light blue dots show the results obtained using heralded single photons. The gray solid line indicates the average classification accuracy over 1000 simulations of training and classification. The gray shaded area represents the variance range of the results obtained from those 1000 simulations.
}
 \label{fig4}
\end{figure}

Figure~\ref{fig4}(a) shows the classification results obtained using a classifier trained with an average sample number of $M = 1.6$ and 200 training data points. Despite significant statistical fluctuations expected at such low photon counts, the cost function consistently decreased during training, and the final classification accuracy reached an impressive 86.1\%. This result highlights not only the robustness of the Data Reuploading architecture but also the effectiveness of the SMO-based optimization under real experimental conditions.

To further assess the classifier’s performance, we conducted classification experiments by varying the number of samples $M$ used for training, while keeping the number of training data points fixed at 200. Due to the low count rate of the heralded single-photon source, we employed a hybrid strategy: a pseudo-single-photon source was used for experiments requiring large photon numbers (e.g., $M > 20$), while the heralded source was used in the photon-starved regime (e.g., $M \leq 20$), where the behavior of true single photons is most relevant. This separation allowed us to systematically evaluate the classifier’s performance across a wide range of sampling conditions.

Figure~\ref{fig4}(b) shows the classification results obtained using classifiers trained on 200 fixed training data points, with the average sample number $M$ varying from approximately 2 to 5000. As shown in the figure, the classification accuracy increased with $M$ and saturated above 90\% for large $M$. Moreover, similar to the simulation results, even with only a few samples and 200 training data points, the classifier achieved good overall accuracy, demonstrating its robustness in the low-photon regime.


\section{Conclusions}
In this study, we experimentally investigated the performance of a single-qubit quantum classifier implemented on a silicon photonic integrated circuit, with a particular focus on the photon-limited regime. Using the Data Reuploading method and layer-wise optimization via the Sequential Minimal Optimization (SMO) algorithm, we demonstrated that effective training is possible even with a severely limited number of photon samples. Heralded single photons, generated through spontaneous four-wave mixing in a silicon waveguide, served as inputs to the circuit. Our evaluation of classification accuracy as a function of both photon sample size and training dataset size revealed that even with an average of approximately two photons per input, accuracies exceeding 90\% can be achieved, given sufficiently large training datasets.
These results provide empirical support for the theoretical prediction that the trade-off between quantum sampling resources and classical training data volume can be strategically balanced. The demonstrated ability to maintain high accuracy with minimal photon usage highlights the practical feasibility of energy-efficient, low-resource quantum machine learning using photonic platforms.

Looking forward, scaling this approach to more complex quantum circuits and multi-qubit systems could further enhance model expressiveness and classification capability. Moreover, integrating classical preprocessing or postprocessing may lead to hybrid architectures that combine the advantages of both classical and quantum computation. Such developments will be essential for realizing scalable, resource-aware quantum machine learning models suitable for real-world applications.
Future work will explore extending the training dataset beyond the current computational limits of 1000 examples, and investigating whether performance degradation at very low photon sample numbers (e.g., $M<10$) can be mitigated through increased data volume or alternative training strategies.

\textbf{\emph{Acknowledgements}}
This work was supported by JST PRESTO Grant No. JPMJPR1864, and JST ERATO Grant No. JPMJER2402, JSPS KAKENHI Grant Number JP24K00559, JST COI-NEXT Grant No. JPMJPF2221, and JST Moonshot R \& D Grant No. JPMJMS2061, the Murata Science Foundation, the Shimazu Science Foundation and the Casio Science Promotion Foundation.

\bibliography{reference}

\end{document}